\documentclass{ws-ijmpd}
\usepackage[super,compress]{cite}
\begin{document}

\markboth{Ivan Zh. Stefanov}
{Mass-spin relation of BHs obained by twin HF QPOs}

%
\catchline{}{}{}{}{}
%

\title{MASS-SPIN RELATION OF BLACK HOLES OBTAINED BY TWIN HIGH-FREQUENCY QUASI-PERIODIC OSCILLATIONS}

\author{IVAN ZHIVKOV STEFANOV}

\address{Technical University of Sofia, \\Faculty of Applied Mathematics and Informatics,\\ Department of Applied Physics,\\8,~Snt.~Kliment Ohridski Blvd., 1000 Sofia, Bulgaria\\
izhivkov@tu-sofia.bg}

\maketitle

\begin{history}
\received{Day Month Year}
\revised{Day Month Year}
\end{history}

\begin{abstract}
The paper studies the uniqueness and the monotonicity of the mass-spin relation of black holes the X-ray power density spectra of which contain twin high-frequency quasi-periodic oscillations in 3:2 ratio.  It is found that for geodesic models the properties of the mass-spin relation are independent of the observed frequencies, i.e. they are independent of the particular object. Some results are valid for all geodesic models. For concreteness two of the most commonly used models are studied here -- the $3:1$ nonlinear epicyclic resonance model and its Keplerian version. At the end, the admissible values of the mass of the black hole as function of the observed upper frequency are plotted.
\end{abstract}


\ccode{PACS numbers:}


\section{Introduction}
Twin high-frequency quasi-periodic oscillations (HF QPOs) observed in the power density spectra of black holes, which are commonly hypothesized to occur in 3:2 ratio \cite{Klis_REVIEW, 32universal}, allow us to measure their spins $a$, if their masses $M$ are known \cite{Abramowicz_GRO_spin_2001, NRM1, NRM2}. If the mass of black hole is not known in advance, a mass-spin relation can be obtained instead \cite{mass_angular_relation_Chi2_2010, Stuchlik_confront, mass_angular_relation_Chi2_2014, moyataAN, StuchlikKolosGRO}. An important question that can be asked in this situation is whether the $M(a)$ function is single-valued and monotonous? If it is not single-valued, then two or more black holes with equal spins but different masses may have produced a give pair of observed HF QPOs. If the function is not monotonous another type of degeneracy occurs -- the observed frequencies may be attributed to two or more black holes which have equal masses but different spins. Since the $M(a)$ function is defined only implicitly, on the one hand, and depends on the observed frequencies, on the other, the answer to this question is not straightforward.

In the current paper we compose the system which defines the $M(a)$ function implicitly and rearrange it to a form which eases the analysis in Section~1. In this section we also plot the $M(a)$ relation for the microquasar GRO~J1655-40 obtained by two of the most commonly used models the $3:1$ nonlinear epicyclic resonance model ($3:1$ NRM) and its Keplerian version ($3:1$ K NRM), as an example. Then, in Section~2 the derivative $M_{,a}(a)$\footnote{Here and, similarly, bellow the comma symbol designates a derivative.} is evaluated. Conclusions about the uniqueness and the monotonicity of the $M(a)$ function that the two models studied here provide is made in Section~3. Diagrams of the bounds on $M$ that can be obtained with these models as function of one of the two observed frequencies are plotted in Section~4. Some general observations are made in the Conclusion.

In this paper all the masses are scaled with the Solar mass, the radii are scaled with the gravitational radius $r_{\rm g}\equiv G M/c^2$,  and the specific angular momentum $a\equiv cJ/G M^2$ is used.
\section{Spin-mass relation}
Of all the models of the 3:2 twin high-frequency QPOs that can be found in the literature the geodesic ones comprise the largest group. Their relative simplicity makes them very attractive. (See \cite{StuchlikKolosGRO} for a list of the most commonly applied geodesic models.) These models attribute the occurrence of observed pair of HF QPOs to the frequencies of epicyclic motion of inhomogeneities in the accretion disk surrounding the black hole, treated as point particles, or to simple linear combinations of them. The frequencies of epicyclic motion of a point particle orbiting about a Kerr black hole, the radial $\nu_{ \rm r}$, the vertical $\nu_{ \rm \theta}$ and the orbital frequency $\nu_{ \rm \phi}$, are given here in the Appendix.

Some of the results obtained here are model independent. For concreteness we use as an example two akin models -- the $3:1$ NRM and the Keplerian version of this model the $3:1$ K NRM \cite{NRM1,NRM2,NRM3}. In the first model the resonance occurs between the radial
epicyclic frequency and the vertical one. These frequencies are functions of the radial coordinate. The resonance occurs on the orbit for which the following ratio $\nu_{ \rm \theta}:\nu_{ \rm r} = 3:1 $  occurs.  According to this model, the observed twin HF QPOs are manifestations of the frequencies $\nu_{ \rm U}=\nu_{ \rm \theta}$ and $\nu_{ \rm L}=\nu_{ \rm \theta}-\nu_{ \rm r}$ which are in 3:2 ratio. In the latter one the vertical epicyclic frequency is replaced by the orbital frequency. In the special case of twin HF QPOs in 3:2 ratio the second model, actually, coincides with the relativistic precession model \cite{RP_1, RP2, RP3} which is by far the most commonly used.

Using the model functions for the two HF QPOs and the observed values, $\nu_{ \rm L}^{ \rm  obs}$ and $\nu_{ \rm U}^{ \rm  obs}$ we can compose the following system
\begin{eqnarray}
&&\nu_{ \rm  L}(a,M,r_{ \rm HF})=\nu_{ \rm L}^{ \rm  obs}, \label{eq_L}\\
&&\nu_{ \rm U}(a,M,r_{ \rm HF})=\nu_{ \rm U}^{ \rm  obs} \label{eq_U}.
\end{eqnarray}
The explicit form of $\nu_{ \rm  L}(a,M,r)$ and $\nu_{ \rm  U}(a,M,r)$ depends on the choice of a model for the HF QPOs and of a metric.

The current study is constrained to the case $\nu_{ \rm U}^{ \rm  obs}:\nu_{ \rm L}^{ \rm  obs}=3:2$ since it is believed that this ratio is a general property of the HF QPOs of black holes \cite{Klis_REVIEW, 32universal}.

It is convenient to express the system (\ref{eq_L})--(\ref{eq_U}) in the following form
\begin{eqnarray}
&&f_1\equiv2\bar{\nu}_{ \rm U}(a,r)-3\bar{\nu}_{ \rm L}(a,r)=0, \label{eq1}\\
&&f_2\equiv \bar{\nu}_{ \rm  U}(a,r) -M\nu_{ \rm U}^{ \rm  obs} =0. \label{eq2}
\end{eqnarray}
Here and bellow the bar symbol designates the part of the expression for the frequencies that is independent of $M$
\begin{eqnarray}
\nu_j(a,M,r)={\bar{\nu}_j(a,r)\over M}, \quad\quad j=U,\,L.
\end{eqnarray}
Equation (\ref{eq1}) is independent of the mass $M$ and allows us to find the position of the radius on which the upper and the lower frequency are in 3:2 ratio as function of the spin.

\begin{figure}
    \includegraphics[width=0.49\textwidth]{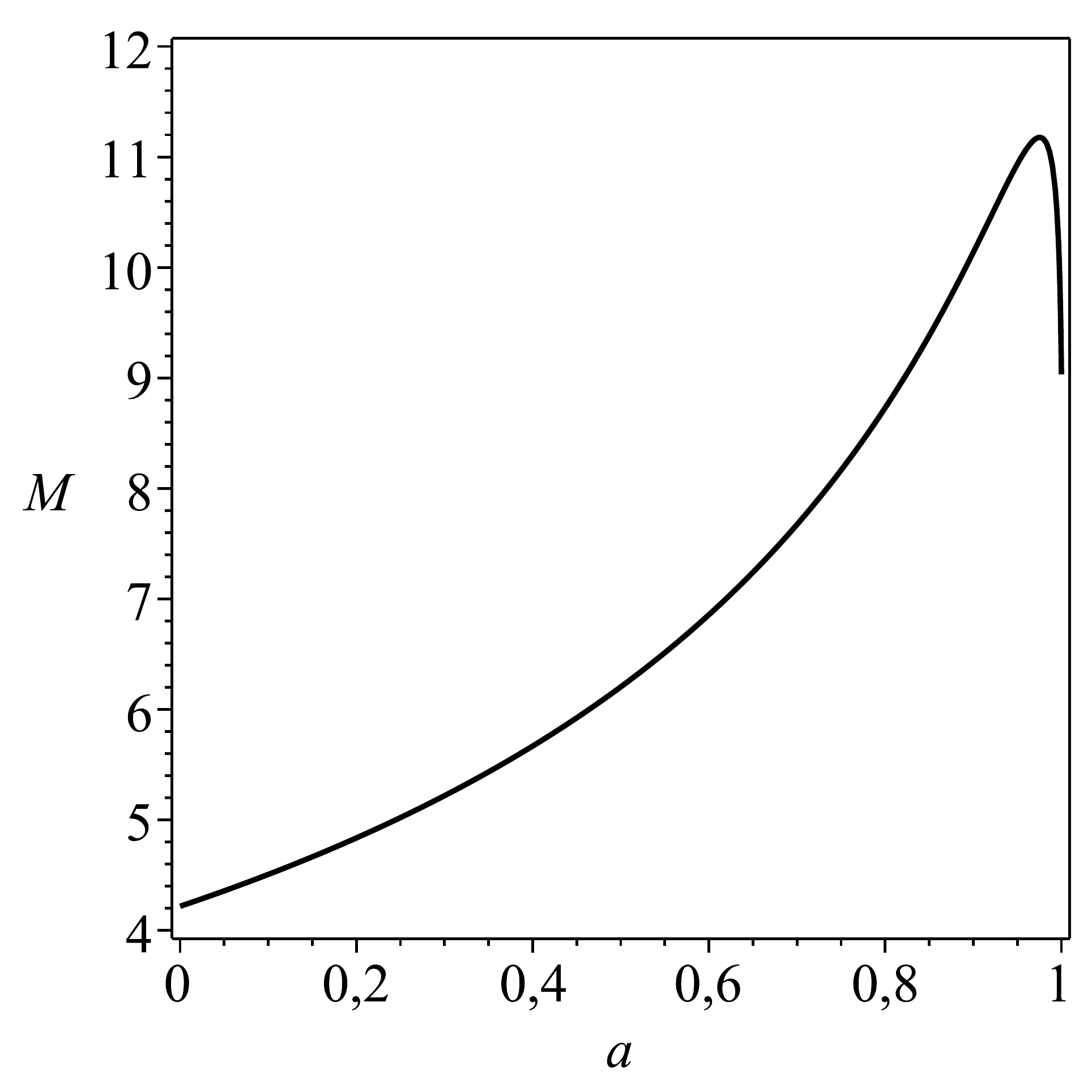}
    \includegraphics[width=0.49\textwidth]{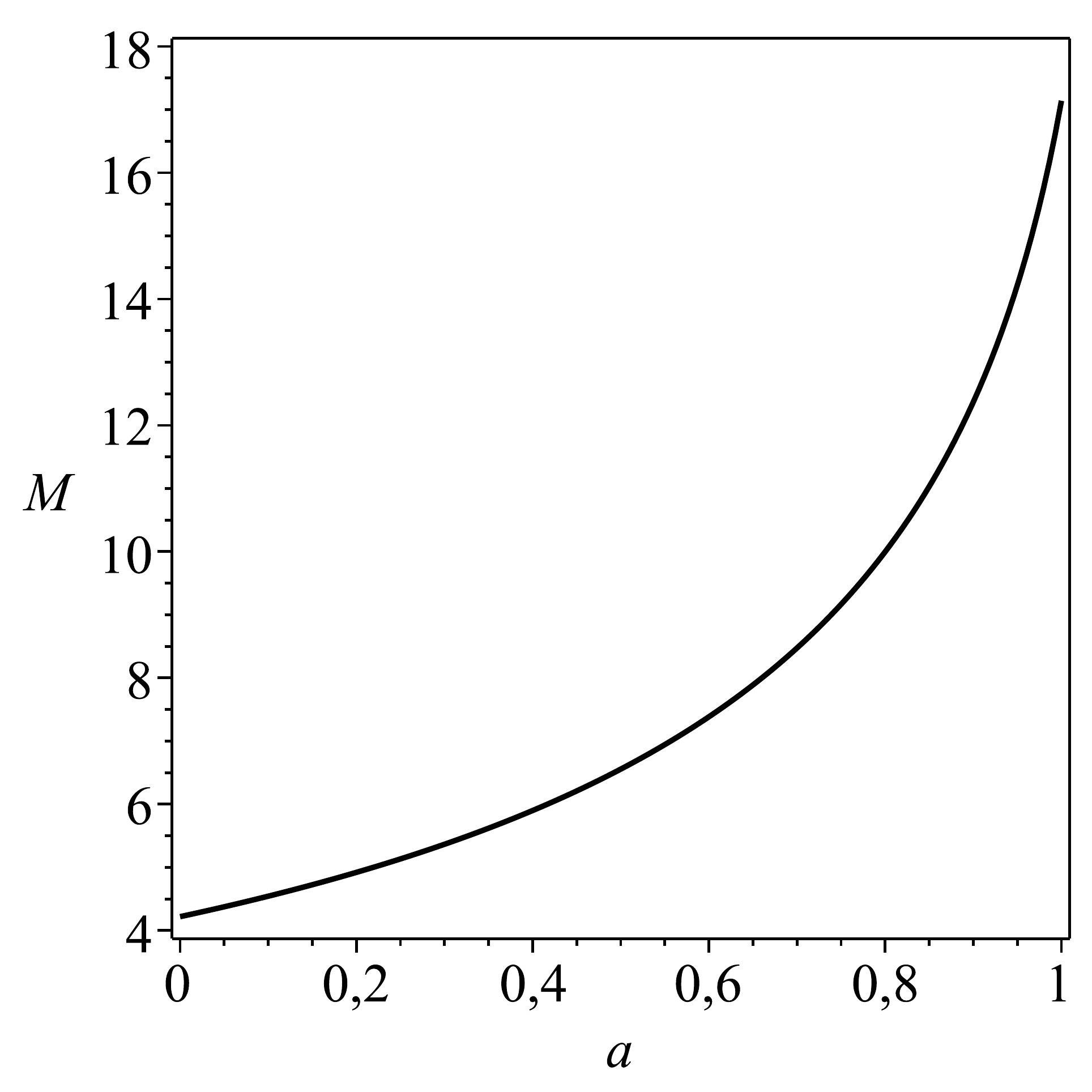}\\
  \caption{The $M(a)$ relation of the microquasar GRO~J1655-40,  obtained by (\ref{eq1})--(\ref{eq2}), for the case of 3:1 NRM model (left panel) and 3:1 K NRM (right panel).}\label{Ma_GRO}
\end{figure}

The system (\ref{eq1})--(\ref{eq2}) has two equations and three parameters $M$, $r$ and $a$. The first two parameters are implicitly defined functions of $a$. It is more convenient to choose $a$ as an independent variable, rather than $M$ or $r$, since the former varies in a closed interval, namely $a\in[0,1]$ for a Kerr black hole.
The $M(a)$ relation obtained by (\ref{eq1})--(\ref{eq2}) is exemplified on Fig.~(\ref{Ma_GRO}) for the special case of the microquasar GRO~J1655-40, for which $\nu_{ \rm U}^{ \rm  obs}=441\pm2 \, {\rm Hz}$ \cite{RP_to_GRO_Motta}. 
As it can be seen from Fig.~(\ref{Ma_GRO}), the $M(a)$ function obtained with the $3:1 K$ NRM model  is monotonous, while in the case of 3:1 NRM it has one maximum.
\section{Derivative of the mass $M_{,a}(a)$}
As the function $M(a)$ is defined implicitly by the system (\ref{eq1})--(\ref{eq2}), its derivative is evaluated with the help of the corresponding Jacobian determinants \cite{MultivariableCalculus1, MultivariableCalculus2}
\begin{eqnarray}
M,a=-\frac{D(f_1,f_2)}{D(a,r)}.\left[\frac{D(f_1,f_2)}{D(M,r)}\right]^{-1}.\\
\end{eqnarray}
The numerator would vanish at the extrema of $M(a)$, if they exist. If the denominator,the Jacobian in the square brackets, has no zeroes, then, according to the implicit function theorem \cite{MultivariableCalculus1, MultivariableCalculus2}, the functions $M(a)$ and $r(a)$ are single-valued. If turning points were present -- points at which two branches of the $M(a)$ function merge, the denominator would vanish at them.

The explicit form of the numerator is
\begin{equation}
\frac{D(f_1,f_2)}{D(a,r)}=3\left(\bar{\nu}_{ \rm L,r}\bar{\nu}_{ \rm U,a}-\bar{\nu}_{ \rm L,a}\bar{\nu}_{ \rm U,r}\right).\label{numer}\\
\end{equation}
The denominator is
\begin{equation}
\frac{D(f_1,f_2)}{D(M,r)}=\nu_{ \rm U}^{ \rm  obs}\left(2\bar{\nu}_{ \rm U,r}-3\bar{\nu}_{ \rm L,r}\right).\label{numer}\\
\end{equation}
The first observation that can be made is that both the numerator and the denominator are independent of $M$. This means that in order to make conclusion about the presence of extrema and turning points in the $a-M$ diagram it is sufficient to study the subspace of the parameters $a$ and $r$. This result is valid for the entire class of geodesic models. The zeroes  of these Jacobins are independent also of the observed frequencies.

The zeroes of the numerator or, respectively the denominator, if any, would be represented by curves in the $a-r$ space. Only those roots of the Jacobian which are within the region $[0,1]\times\left[\right.r_{\rm ISCO},\infty\left.\right)$, where ``$\times$'' designates the Cartesian product of the intervals of admissible values of $a$ and $r$, respectively, may have physical significance. Here and bellow ISCO stands for the innermost stable circular orbit, which here is accepted to be the inner boundary of the accretion disk.
\section{Extrema of $M(a)$}\label{extrema}
For the 3:1 K NRM model, neither the numerator, nor the denominator of $M_{,a}(a)$ has zeroes, as the numerical study reveals. As it was mentioned in Section~1, the $M(a)$ function obtained with this model is monotonous.

\begin{figure}
\center
  \includegraphics[width=0.49\textwidth]{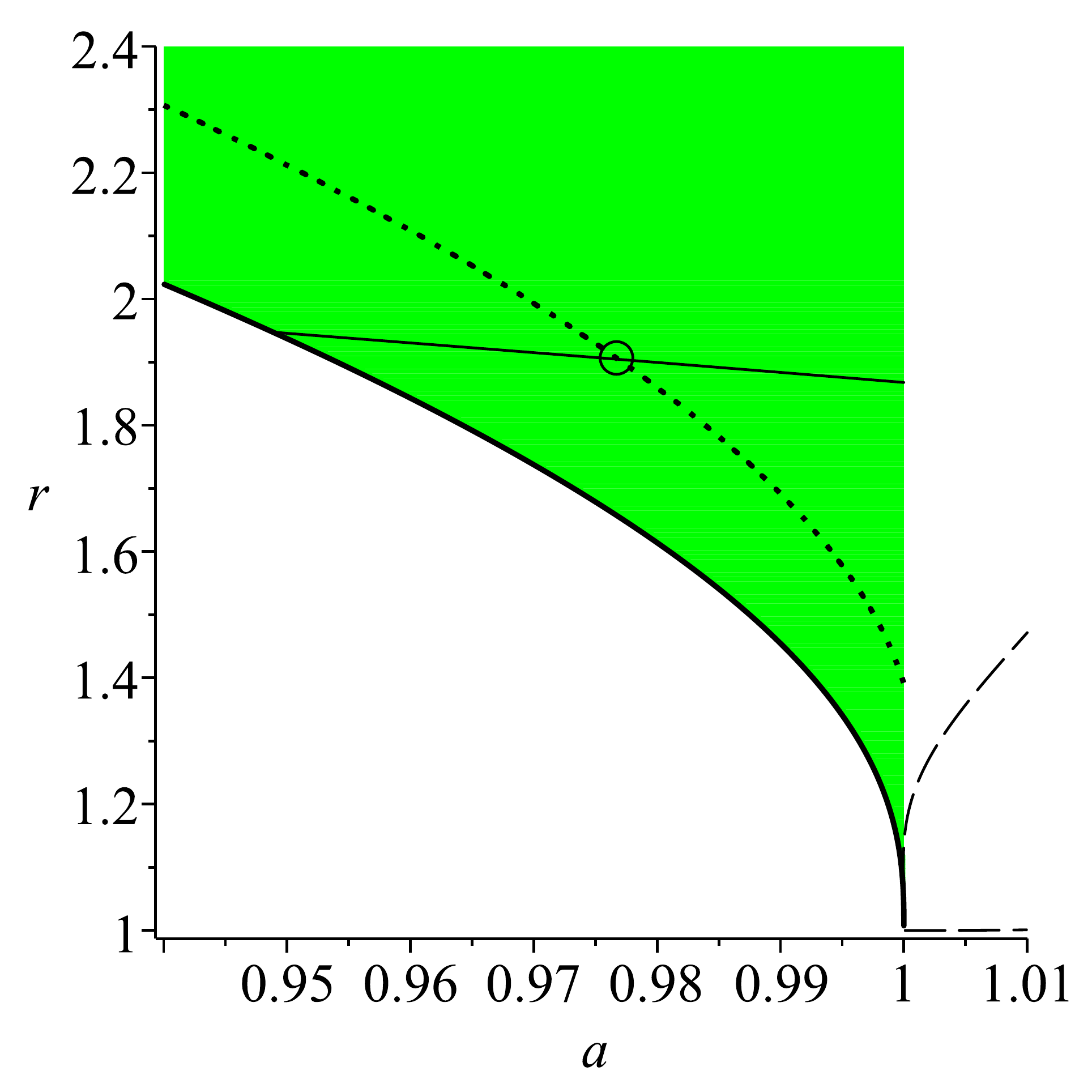}\\
  \caption{Characteristic radii: ISCO (solid), zeroes of the denominator (thin solid) and the numerator (thin dashed) of $M_{,a}(a)$ and radius of the resonant orbit (dotted)for the case of 3:1 NRM.}\label{radii}
\end{figure}

The zeroes of the denominator and the numerator of $M_{,a}(a)$ for the case of 3:1 NRM are represented by thin solid and thin dashed line on Fig.\ref{radii}. The physically significant region $[0,1]\times\left[\right.r_{\rm ISCO},\infty\left.\right)$ is colored in green. The thick solid line represents the radius of the ISCO as a function of $a$.
The zeroes of the denominator are outside of the physically interesting region colored in  green on Fig.\ref{radii}, so the $M_{,a}(a)$ function has one branch, i. e. it is single-valued. The denominator, however, has one zero in the green region. It would be significant, only if it is also a root of the system (\ref{eq1})--(\ref{eq2}), i.e. only if it intersects the $r_{3:2}$ radius. As it can be seen on Fig. \ref{radii}, they do intersect and the intersection point is designated by a circle. The crossing point allows us to obtain $a_{\rm extr}$, the value of the spin for which $M(a)$ has an extremum, a maximum in particular. Its value is $a_{\rm extr}=0.977$. This result is independent of the values of the observed frequencies and is, hence, valid for all objects for which the 3:1 NRM is applicable.

\section{Conclusion}
The current study focuses on properties of the mass-spin relation of black holes the X-ray power density spectra of which contain twin high-frequency quasi-periodic oscillations in 3:2 ratio. The $M(a)$ function is defined implicitly. It was found that the presence or absence of extrema and turning points of the $M(a)$ function and their positions are independent of the values of the observed twin HF QPOs, $\nu_{ \rm L}^{ \rm  obs}$ and $\nu_{ \rm U}^{ \rm  obs}$. Hence the monotonicity and the uniqueness of this function  is independent of the particular object but is, rather, an intrinsic property of a given geodesic model.

The Jacobians that appear in the numerator and the denominator of the derivative of $M(a)$ have been evaluated. They are independent of $M$. In order to make conclusions about the presence of extrema or branches of the $M(a)$  function that a given geodesic model provides one has to study the subspace of the parameters $a$ and $r$.

Two particular models have been studied as an example -- 3:1 NRM and the 3:1 K NRM. For the first model $M(a)$ is monotonous and single-valued. For the second model, $M(a)$ has one extremum at  $a=a_{\rm extr}=0.977$. The extremal value of the mass $M\left(a_{\rm extr}\right)$ is inversely proportional to the upper (or, alternatively, the lower) frequency. 

\section*{Acknowledgements}
I.S. would like to thank his wife for the support, Dr. Sava Donkov and Dr. Radostina Tasheva for the numerous discussions on the subject. The research is partially supported by the the Bulgarian National Science Fund under Grant N 12/11 from 20 December 2017.

\appendix
\section{Epicyclic frequencies}\label{app_freqs}
The explicit form of the orbital frequency $\nu_{\rm \phi}$ and the two epicyclic frequencies -- the radial $\nu_r$ and the vertical $\nu_{\theta}$ -- for the Kerr black hole \cite{AlievGaltsov1, AlievGaltsov2,AlievKerr}
\begin{eqnarray}
\nu_{\rm \phi} &=&\left({c^3\over 2\pi GM}\right)\frac{ 1}{ r^{3/2} \pm a}, \label{nu_phi}\\
\nu_{r}^2 &=& \nu_{\rm \phi}^2\, \left( 1-\frac{6 }{r} -\frac{3
a^2}{r^2} \pm {8 a\over r^{3/2}}\right),\label{nu_r}\\
\nu_{\theta}^2&=& \nu_{\rm \phi}^2\, \left(1
+\frac{3 a^2}{r^2} \mp {4 a \over r^{3/2}} \right)\label{nu_theta}.
\end{eqnarray}
The upper (lower) sign corresponds to prograde (retrograde) direction of rotation of the hot spot.



\begin{thebibliography}{00}    
\bibitem{Abramowicz_GRO_spin_2001} Abramowicz M. A. and Kluzniak W., 2001, A\&A 374, L19 -- L20.	
\bibitem{NRM1} Abramowicz M. A., Kluzniak W., AIP Conference Proceedings \textbf{714}, 21(2004), arXiv:astro-ph/0312396.
\bibitem{NRM2} Abramowicz M. A., Kluzniak W., Stuchlik Z., G. T\"{o}r\"{o}k,  arXiv:astro-ph/0401464.
\bibitem{NRM3} Torok G., Abramowicz M. A., Kluzniak W., Stuchlik Z., 2006, AIP Conf. Proc., 861, 786, arXiv:astro-ph/0603847.
\bibitem{AlievGaltsov1} Aliev A. N. and Gal'tsov D. V., 1981, Gen. Relat. Gravit. \textbf{13}, 899.
\bibitem{AlievGaltsov2} Aliev A. N., Gal'tsov D. V. and Petukhov V. I., 1986., Astr. Space Sci. \textbf{124}, 137.
\bibitem{AlievKerr} Aliev A. N., Esmer G. D., Talazan P.,
    2013, Class. Quantum Grav. \textbf{30}, 045010.
\bibitem{MultivariableCalculus1}Botelho F. S., 2018, \emph{Real Analysis and Applications}, Springer International Publishing AG part of Springer Nature, Cham, Switzerland.
\bibitem{MultivariableCalculus2} Ilyin  V.A., Sadovnichiy  V.A., Sendov  B.Kh., 2013, \emph{Matematicheskiy analiz.  V  2chastyakh.  Ch. 1}  [ \emph{ Mathematical  analysis.   In  2  parts.  Part  1 } ], YuraytPubl, Moscow (in Russian).
\bibitem{RP3} Merloni A., Vietri M., Stella L., Bini D., 1999, MNRAS, 304, 155.
\bibitem{RP_to_GRO_Motta} Motta S. E., Belloni T. M., Stella L., Munoz-Darias T., Fender R., 2014a, MNRAS  437, 2554.
\bibitem{moyataAN} Stefanov I. Zh., 2016, Astron. Nachr. 337, 246.
\bibitem{RP_1} Stella L., Vietri M., 1998, ApJ, 492, L59.
\bibitem{RP2} Stella L., Vietri M., Morsink S., 1999, ApJ, 534,
\bibitem{StuchlikKolosGRO}Stuchl\'{\i}k Z., Kolo\v{s} M., 2016, ApJ 825, 9.
\bibitem{NRM3} T\"{o}r\"{o}k G., Abramowicz M. A., Kluzniak W., Stuchl\'{i}k Z., 2006, AIP Conf. Proc., 861, 786, arXiv:astro-ph/0603847.
\bibitem{mass_angular_relation_Chi2_2010} T\"{o}r\"{o}k G.,  Bakala P., Sr\'{a}mkov\'{a} E., Stuchl\'{\i}k Z., Urbanec M.: 2010, ApJ 714, 748.
\bibitem{Stuchlik_confront} T\"{o}r\"{o}k G., Kotrlov\'{a} A., \v{S}r\'{a}mkov\'{a} E., Stuchl\'{i}k Z., 2011, A\&A, 531, A59.
\bibitem{mass_angular_relation_Chi2_2014} T\"{o}r\"{o}k G., Bakala P.,\v{S}r\'{a}mkov\'{a} E., Stuchl\'{i}k Z., Urbanec M., Goluchov\'{a} K.,  2012, ApJ 760, 138.
\bibitem{Klis_REVIEW} van der Klis M., 2006, in Compact Stellar X-Ray Sources, ed. W. H. G., Lewin \& M. van der Klis (Cambridge: Cambridge Univ. Press), 39, arXiv:astro-ph/0410551.
\bibitem{32universal} Zhou X. L., Yuan W., Pan H. W., Liu Z., 2015, ApJ 798, 4.
\end{thebibliography}
\end{document}